\documentclass[12pt]{article}
\usepackage{graphicx}
\usepackage[utf8]{inputenc}
\usepackage[english
]{babel}
\def\baselinestretch{1.5}
\begin{document}

\begin{center} {\Large \bf
Quantum suprematism picture of Triada of  Malevich's squares for spin states
and the parametric oscillator evolution in the probability representation of quantum mechanics 
}
\end{center}

\bigskip

\begin{center} {\bf V. N. Chernega$^1$, O. V. Man'ko$^{1,2}$, V. I. Man'ko$^{1,3,4}$}
\end{center}

\medskip

\begin{center}
$^1$ - {\it Lebedev Physical Institute, Russian Academy of Sciences\\
Leninskii Prospect 53, Moscow 119991, Russia}\\
$^2$ - {\it Bauman Moscow State Technical University\\
The 2nd Baumanskaya Str. 5, Moscow 105005, Russia}\\
$^3$ - {\it Moscow Institute of Physics and Technology (State University)\\
Institutskii per. 9, Dolgoprudnyi, Moscow Region 141700, Russia}\\
$^4$ - {\it Tomsk State University, Department of Physics\\
Lenin Avenue 36, Tomsk 634050, Russia}\\
Corresponding author e-mail: omanko@sci.lebedev.ru

\end{center}

\smallskip

\smallskip

\begin{abstract}
Review of tomographic probability representation of quantum states is presented both for oscillator systems with continious variables and spin--systems with discrete variables. New entropy--information inequalities are obtained for Franck--Condon factors. Density matrices of qudit states are expressed in terms of probabilities of artificial qubits as well as the quantum suprematism approach to geometry of these states using the triadas of Malevich squares is developed. Examples of qubits, qutrits and ququarts are considered.
\end{abstract}
Keywords: qubit, qudit, tomographic  probability representation, entropy--information inequalities,  Franck--Condon factors, quantum suprematism approach.

\section{Introduction}
In quantum mechanics the system states are associated either with vectors $|\psi\rangle$ in Hilbert spaces \cite{Dirac} as well as with wave functions (of position) $\psi(q)$ \cite{Schr26} (for pure states) or with density matrices \cite{Landau27,vonNeumann27} (for mixed states). The physical observables are described by Hermitian operators acting in the Hilbert space.

This formalism is radically different from the formalism of classical mechanics where the states of a particle are associated with its position $q$ and momentum $p$ (if there is no fluctuations of these classical observables) or with  probability distributions $f(q,p)$ (if there present fluctuations of these observables). We wish to create equivalent quantum formalism which is closer to classical formalism yielding the result obtained by Wigner \cite{Wigner32} who introduced the Wigner function $W(q,p)$. This function is similar to classical probability distribution $f(q,p)$ but the function can take negative values and in view of this it is not probability distribution. Analogous functions on phase--space are Glauber--Sudarshan $P$-function \cite{Glauber63,Sud63,Cah},  Husimi--Kano $Q$--function \cite{Husimi,Kano56} connected with the Wigner function by integral transform. These functions are called quasidistributions. For electron spin the pure state is associated either with vector $|\psi\rangle$ in two--dimensional Hilbert space (spinor) or for mixed state with $2\times2$ - density matrix (density operator $\hat\rho$) acting in this Hilbert space. In the work of Stratonovich \cite{Stratonovich57} the formalism of analog of the quasidistributions for spin-1/2 particle was introduced. But in classical mechanics or classical statistical mechanics namely probability distributions are used to describe the system states. The probability representation for states of quantum systems with continious variables was introduced in \cite{TombesPLA96,TombesiFoundPhyus96} and discussed in \cite{JRLR1997}. This representation called tomographic probability representation is based on the known tool to reconstruct the state Wigner function in quantum--optical experiments \cite{Raymer93} using relation \cite{BerBer,VogRis} of this function with measurable optical tomogram $w(X|\theta)$, which is fair conditional probability distribution \cite{BentyYacolb} of variable $X$ called in quantum optics homodyne quadrature,  depending on local oscillator phase $\theta$, by means of integral Radon transform \cite{Radon1917}. It turned out that the spin states also can be associated with fair probability distributions called spin--tomograms \cite{DodPLA,OlgaJETP,Bregence,Weigert1,Weigert2,PainiDariano}. Since the quantum states in the probability representation are identified with probability distributions all the instruments and results of the classical probability theory were  applied to get new results in quantum mechanics, quantum optics and quantum information (see e.g. reviews \cite{IbortPhysScr2009,IbortPhysScr2015,MendesJRusLasRes}). The aim of this work is to review the probability representation for quantum states on example of parametric oscillator and spin--states. Using the known physical meaning of Frank--Condon factors in molecular spectroscopy as probability distributions we describe some new entropy--information inequalities studied in quantum mechanics \cite{MAVI} for these factors. These inequalities can be useful in study of vibronic spectra of polyatomic molecules discussed e.g. in \cite{MalkinMolSP1076,MalkinMolSpec1977,HuberHu,ZhabrakJRusLasRes}. Also for spin--states we present new formulas for the density matrices expressed in terms only of probabilities of spin projections \cite{MarmoJPA2016,CherJRLR1,CherJRLR2,CherJRLR32017} and provide the quantum suprematism picture of the spin--states where triangle geometry of the states is associated with triadas of Malevich's squares \cite{CherJRLR1,CherJRLR2,CherJRLR32017}. The review of suprematism approach in art is given in \cite{Malevich}. 

The paper is organised as follows.

In Sec. 2 we review the optical and symplectic tomographic probability distributions on examples of oscillator states including parametric oscillator states. In Sec. 3 we discuss new entropy--information inequalities for Franc--Condon factors of diatomic and polyatomic molecules. In Sec. 4 we review the spin--tomography for qudit states. In Sec. 5 we present explicit expressions of spin--state density matrices for qubit and qutrit states in terms of probabilities of spin--1/2 projections on perpendicular directions and discuss the expressions of such density matrices for states of arbitrary spins. In Sec. 6 we review the quantum suprematism representation of qudit states. The conclusions and prospectives are given in Sec. 7.

\section{Tomographic probability distributions of states with continious variables}
We consider the systems like oscillator with wave function $\psi(x), -\infty\leq x\leq \infty$. The Wigner function of the system is given by relation ($\hbar=m=1$)
\begin{equation}\label{eq.1.1}
W(q,p)=\int\psi(q+u/2)\psi^\ast(q-u/2)e^{-i p u}d u.
\end{equation}
The Wigner function is real function and for normalized states, i.e. $\int|\psi(x)|^2d x=1$, it is normalized
\begin{equation}\label{eq.1.2}
\frac{1}{2\pi}\int W(q,p)d q d p=1.
\end{equation}
The density matrix of the state $\rho_{\psi}(x,x')=\psi(x)\psi^\ast(x')$ is given by the transform
\begin{equation}\label{eq.1.3}
\rho_{\psi}(x,x')=\frac{1}{2\pi}\int W(\frac{x+x'}{2},p)e^{ip(x-x')} dp.
\end{equation}
The tomographic probability distribution called symplectic tomogram $w(X|\mu,\nu)$ where $-\infty\leq X,\mu,\nu\leq\infty$ is defined by the Radon transform of the Wigner function \cite{TombesiSemOpt}
\begin{equation}\label{eq.1.4}
w(X|\mu,\nu)=\int W(q,p)\delta(X-\mu q-\nu p)\frac{d q d q}{2\pi}.
\end{equation}
The function is nonnegative and normalized for arbitrary parameters $\mu$ and $\nu$, i.e.
\begin{equation}\label{eq.1.5}
\int w(X|\mu,\nu)d X=1.
\end{equation}
It can be easily checked using (\ref{eq.1.1}) that the symplectic tomogram of pure state with wave function $\psi(y)$ is given by the fractional Fourier transform of the wave function \cite{MendesPLA}
\begin{equation}\label{eq.1.6}
w(X|\mu,\nu)=\frac{1}{2\pi|\nu|}\,|\int\psi(y)exp\left(\frac{i\mu y^2}{2\nu}-\frac{i X y}{\nu}\right)d y|^2.
\end{equation}
For the parameters $\mu=\cos\theta$, $\nu=\sin\theta$ the symplectic tomogram coincides with probability distribution called optical tomogram $w(X|\theta)$ which reads
\begin{equation}\label{eq.1.7}
w(X|\theta)=\int W(q,p)\delta(X-q\cos\theta-p\sin\theta)\frac{d q d p}{2\pi}.
\end{equation}
For pure state the optical tomogram has the following form
\begin{equation}\label{eq.1.8}
w(X|\theta)=|\int\psi(y)\frac{\exp\left(\frac{i\cot \theta}{2}(X^2+y^2)-\frac{i X y}{\sin\theta}\right)}{\sqrt{2\pi i\sin\theta}}d y|^2.
\end{equation}
The symplectic tomogram determines the Wigner function
\begin{equation}\label{eq.1.9}
W(q,p)=\frac{1}{2\pi}\int w(X|\mu,\nu)\exp(i(X-\mu q-\nu p))d X d \mu d \nu
\end{equation}
and also the density operator $\hat\rho$ of the state with the Wigner function, namely \cite{TombesiFoundPhyus96}
\begin{equation}\label{eq.1.10}
\hat\rho=\frac{1}{2\pi}\int w(X|\mu,\nu)\exp(i(X\hat1-\mu\hat q-\nu \hat p))d X d \mu d \nu.
\end{equation}
Here $\hat q$ and $\hat p$ are position and momentum operators.

\section{Frank-Condon factors on example of parametric oscillator}
For oscillator with varying frequency $\omega(t)$ and the Hamiltonian
\begin{equation}\label{eq.2.1}
\hat H=\frac{\hat p^2}{2}+\frac{\omega^2(t)\hat q^2}{2}, \quad \omega(0)=1
\end{equation}
the solutions of Schr\"odinger equation can be obtained using the integrals of motion $\hat A(t),\,\hat A^\dagger(t)$ found in \cite{MalkinMankoPhysLet1966}, namely
\begin{equation}\label{eq.2.2}
\hat A(t)= \frac{i}{\sqrt 2}\left(\epsilon(t)\hat p-\dot\epsilon(t)\hat q\right),
\end{equation}
where
\begin{equation}\label{eq.2.3}
\ddot\epsilon(t)+\omega^2(t)\epsilon(t)=0
\end{equation}
and $\epsilon(0)=1,\,\dot\epsilon(0)=i,\,\hat A(0)=\frac{\hat q+i\hat p}{\sqrt 2}$. The product of the invariants $\hat A^\dagger(t)\hat A(t)$ equals to Ermakov invariant \cite{Ermakov}. Commutation relations of the integrals of motion (\ref{eq.2.2}) are $[\hat A(t),\hat A^\dagger(t)]=1$. The ground--like state $\psi_0(x,t)$ of the parametric oscillator satisfying the condition $\hat A(t)\psi_0(x,t)=0$ reads
\begin{equation}\label{eq.2.4}
\psi_0(x,t)=\pi^{-1/4}(\epsilon(t))^{-1/2}\exp\left[\frac{i\dot\epsilon(t)x^2}{2\epsilon(t)}\right].
\end{equation}
Fock states $\psi_n(x,t),\,n=0,1,2,\ldots,$ solutions of Schr\"odinger equation are expressed in terms of the integrals of motion and ground--like state as
\begin{equation}\label{eq.2.5}
\psi_n(x,t)=\frac{(\hat A^\dagger(t))^n}{\sqrt{ n!}}\psi_0(x,t),\quad n=0,1,2,\ldots
\end{equation}
They are expressed in terms of Hermite polynomials (see,e.g.\cite{183}). The Frank-Condon factors are probabilities given by the scalar--product
\begin{equation}\label{eq.2.6}
P_m(n,t)=|\langle m,0|n,t\rangle |^2
\end{equation}
where $|m,0\rangle$ is the state at time $t=0$. The Frank--Condon factor is related to vibronic structure of electronic line of two--atomic molecules \cite{MalkinMolSP1076,MalkinMolSpec1977,Huh1,Huh2,Huh3}.

The tomographic probability distribution (\ref{eq.1.6}) of the ground-like state (\ref{eq.2.4}) has the form of normal distribution
\begin{equation}\label{eq.2.6.a}
w_0(X|\mu,\nu)=\frac{1}{\sqrt{2\pi\sigma^2}}\exp\left(-\frac{1}{2\sigma^2}(X-\bar X)^2\right),
\end{equation}
where $\bar X=0$ and
\begin{eqnarray*}\sigma^2=\frac{|\epsilon|^2}{2}\mu^2+\frac{|\dot\epsilon|^2}{2}\nu^2+\mu\nu(|\dot\epsilon\epsilon^\ast|^2-1)^{1/2}.\end{eqnarray*}
One can obtain a new result characterising the properties of the Franck--Condon factors (\ref{eq.2.6}). These factors provide the probability distribution. In view of this the general properties of the probability distributions are present in Franck--Condon factors. For example, Shannon entropy associated with the probability distribution and the inequalities known for this entropy are valid for the factors. If the probability distribution is a joint probability distribution of two random variables $P(a,b)$ there exists inequality for Shannon entropy of the system $H(1,2)=-\sum_a\sum_b P(a,b)\ln P(a,b)$ and two entropies corresponding to marginal distributions 
\[H(1)=-\sum_b\left(\sum_a P(a,b)\right)\ln\left(\sum_a P(a,b)\right)\] 
and 
\[H(2)=-\sum_a\left(\sum_b P(a,b)\right)\ln\left(\sum_b P(a,b)\right)\] 
which means the nonnegativity of the Shannon information 
\[I=H(1)+H(2)-H(1,2)\geq0.\]
If the random variables are not correlated, i.e. $P(a,b)=\Pi(a){\cal P}(b)$ the information $I=0$. The stronger are correlations the larger is the Shannon information. Using the partition tool  \cite{SeilovJRLR,SeilovJRLR1} to present the probability distribution  (\ref{eq.2.6}) as the joint probability distribution we can apply the properties of Shannon information to the Franck--Condon factors. 

We present new entropic inequality for the Franck--Condon factors (\ref{eq.2.6}) of the form

\[-\sum_{n=0}^\infty|\langle m,0|n,t\rangle|^2\ln|\langle m,0|n,t\rangle|^2\leq-\sum_{k=0}^\infty\left[\left(\sum_{j=0}^1|\langle m,0|2k+j,t\rangle|^2\right)\ln\left(\sum_{j=0}^1|\langle m,0|2k+j,t\rangle|^2\right)\right]\]
\begin{eqnarray}
&&-\sum_{j=0}^1\left[\left(\sum_{k=0}^\infty|\langle m,0|2k+j,t\rangle|^2\right)\ln\left(\sum_{k=0}^\infty|\langle m,0|2k+j,t\rangle|^2\right)\right].\nonumber\\
&&\label{eq.2.7}
\end{eqnarray}
The difference of the right--hand side and left--hand side (\ref{eq.2.7}) is Shannon \cite{Shannon} information.
It is nonnegative. The entropic inequality is characterizing the correlations in the vibronic structure of the electronic line in diatomic molecules. This inequality can be checked experimentally. It is worthy to note that the relation and analogies of the molecular spectroscopy method and quantum information technique were studied in \cite{Huh1,Huh2,Huh3}. One can obtain other entropy--information inequalities for the Franck--Condon factors of diatomic and polyatomic molecules using the tool of mapping the integers $s=1,2,3,\ldots,N=\prod_{k=1}^M n_k$ onto sets of of integers $j_1,j_2,\ldots,j_M$ where $ j_1=1,2,,\ldots,n_1,\,j_2=1,2,\ldots,n_2,\ldots,j_n=1,2,\ldots,n_M$ \cite{SeilovJRLR,SeilovJRLR1}. This tool provides possibility to consider Franck-Condon factors as joint probability distributions. These distributions can be mapped onto the joint probability distributions of system with $M$ artificial subsystems. Then the known in probability theory entropy--information inequalities are obviously valid for Franck--Condon factors. The correlations of Franck--Condon factors reflect the correlations of line intensities in vibronic structure of the electronic lines in the molecular spectra. The Franck--Condon factors can be also related to tomographic--probability distribution of quantum states of molecules \cite{ZhabrakJRusLasRes}. In this case the known probability properties of the quantum tomograms provide possibility to obtain new relations for the Franck--Condon factors. It is worthy to note that oscillator with time--dependent parameters describing the dissipation in framework of using Ermakov quadratic  invariant \cite{Ermakov,LewisReisenfeld} was studied in \cite{Shuch1,Shuch2,Shuch3,ShuchCastanos}.

\section{Spin--tomography}
In \cite{DodPLA,OlgaJETP,Bregence} the tomographic probability distribution $w(m|\tilde n)$ was introduced for spin $j$-states and expressed in terms of density matrix $\rho$ and its unitary transform $u$ as diagonal element of the product of three matrices, i.e.
\begin{equation}\label{eq.3.1}
w(m|\vec n)=\left( u(\vec n)\rho u^\dagger(\vec n)\right)_{m m}.
\end{equation}
Here $m=-j,-j+1,\ldots,j-1,j$ is spin--projection on the direction determined by unit vector $\vec n=(\sin\theta\cos\phi,\sin\theta\sin\phi,\cos\theta)$. The unitary matrix $u(\vec n)$ is the matrix of irreducible representation of rotation group corresponding to spin $j$. It depends of three Euler angles $(\psi,\theta,\phi)$ where $\psi=0$. The probability distribution $w(m|\vec n)$ called spin--tomogram determines the density matrix $\rho$. It is normalized
\begin{equation}\label{eq.3.2}
\sum_{m=-j}^j w(m|\vec n)=1
\end{equation}
for arbitrary direction $\vec n$. The density matrix of spin state $\rho=\rho^\dagger$, $\mbox{Tr}\rho=1$, $\rho\geq 0$ depends on $(2j+1)^2-1$ real parameters. In view of this the minimal number of different unit vectors $\vec n$ which is sufficient to express the density matrix in terms of the tomographic probabilities equals also to $(2j+1)^2-1$. For $j=1/2$ it is sufficient to use three such vectors, e.g. to take vectors coinciding with $\vec x,\,\vec y,\,\vec z$ axes, respectively. Let us introduce the notation $w(m=+ \frac{1}{2}|\vec x)=p_1$, $w(m=+ \frac{1}{2}|\vec y)=p_2$, $w(m=+ \frac{1}{2}|\vec z)=p_3$. As it was pointed out \cite{MarmoJPA2016,CherJRLR1,CherJRLR2,CherJRLR32017} the density $2\times2$-matrix $\rho$ is expressed in terms of these three probabilities $p_1,p_2,p_3$ to have positive spin projection $+\frac{1}{2}$ on the axes $\vec x,\vec y,\vec z$, respectively in the form
\begin{equation}\label{eq.3.3}
\rho=
\left(\begin{array}{cc}
p_3&p_1-i p_2-\frac{1-i}{2}\\
p_1+i p_2-\frac{1+i}{2} & 1-p_3\\
\end{array}\right).
\end{equation}
The tomogram (\ref{eq.3.1}) of the spin-$1/2$ state $w(+\frac{1}{2}|\vec n)$ is expressed in terms of the probabilities $p_1,p_2,p_3$ described by the vector $\vec p=(p_1,p_2,p_3)$ and vector $\vec n_0=(1/2,1/2,1/2)$ as the scalar product
\begin{equation}\label{eq.3.4}
w(m|\vec n)=\vec n(\vec p-\vec p_0)+1/2.
\end{equation}
Thus the state of the spin-$1/2$ is identified with three probability distributions determined by three probability vectors  $\vec {\cal P}_1=(p_1,1-p_1),$ $\vec {\cal P}_2=(p_2,1-p_2)$ and $\vec {\cal P}_3=(p_3,1-p_3),$ $1-p_k,\, k=1,2,3$ are the probabilities to have spin projection $m=-1/2$ on the axes $\vec x,\,\vec y,\,\vec z$, respectively. The probabilities must satisfy the quantum relation corresponding to nonnegativity of the density matrix (\ref{eq.3.3}) \cite{MarmoJPA2016}
\begin{equation}\label{eq.3.5}
\sum_{k=1}^3(p_k-\frac{1}{2})^2\leq \frac{1}{4}.
\end{equation}
One can measure in any spin--$1/2$ state the probabilities of spin--projection $m=+1/2$ onto three perpendicular directions. These probabilities must satisfy the above inequality. It was pointed out in \cite{MarmoJPA2016} that this inequality can be checked in experiments with superconducting circuits where analogs of spin states are realized in Josephson junction devices. Different aspects of statistical properties of Josephson junction were discussed in \cite{Vaxjo,physicascr2013,kiktenko}.

\section{Probability representation of arbitrary qudit states}
In \cite{CherJRLR2} the matrix elements of density matrix of qutrit state (spin--$1$ system) were expressed in terms of probabilities of positive spin--$1/2$ projections for three artificial qubits. We rederive this result in the form which can be used to generalize the expression for the density matrix of qutrit and obtain the expression of density matrix of arbitrary qudit state in terms of probabilities associated with the artificial qubits  (spin--$1/2$ projections). Let us consider the $3\times3$-density matrix of a qutrit state and construct two density $4\times4$ matrices of the qudit states with specific zero columns and rows. The tool we use is to start from qutrit density matrix $\rho$
\begin{equation}\label{eq.4.1}
\rho=\left(\begin{array}{ccc}
\rho_{11}&\rho_{12}&\rho_{13}\\
\rho_{21}&\rho_{22}&\rho_{23}\\
\rho_{31}&\rho_{32}&\rho_{33}\\
\end{array}\right).
\end{equation}
and construct the $4\times4$-matrices
\begin{equation}\label{eq.4.2}
\rho(1)=\left(\begin{array}{cccc}
\rho_{11}&\rho_{12}&\rho_{13}&0\\
\rho_{21}&\rho_{22}&\rho_{23}&0\\
\rho_{31}&\rho_{32}&\rho_{33}&0\\
0&0&0&0\\
\end{array}\right),\quad
\rho(2)=\left(\begin{array}{cccc}
0&0&0&0\\
0&\rho_{11}&\rho_{12}&\rho_{13}\\
0&\rho_{21}&\rho_{22}&\rho_{23}\\
0&\rho_{31}&\rho_{32}&\rho_{33}\\
\end{array}\right).
\end{equation}
Interpretating these matrices as the density matrices of two qubit states we construct using the partial tracing procedure four density matrices of the artificial qubit states related to matrix $\rho(1)$, i.e.
\begin{equation}\label{eq.4.3}
R(1)=\left(\begin{array}{cc}
\rho_{11}+\rho_{22}&\rho_{13}\\
\rho_{31}&\rho_{33}\\
\end{array}\right),\quad R(2)=\left(\begin{array}{cc}
\rho_{11}+\rho_{33}&\rho_{12}\\
\rho_{21}&\rho_{22}\\
\end{array}\right),
\end{equation}
and matrix $\rho(2)$
\begin{equation}\label{eq.4.4}
R(3)=\left(\begin{array}{cc}
\rho_{11}&\rho_{13}\\
\rho_{31}&\rho_{33}+\rho_{22}\\
\end{array}\right),\quad R(4)=\left(\begin{array}{cc}
\rho_{22}&\rho_{23}\\
\rho_{32}&\rho_{11}+\rho_{33}\\
\end{array}\right),
\end{equation}
Since these matrices are the density matrices of qubit states one can present these matrices using the probabilities of positive spin--$1/2$ projections, namely
\begin{equation}\label{eq.4.5}
R(k)=\left(\begin{array}{cc}
p_3^{(k)}&(p_1^{(k)}-\frac{1}{2})-i(p_2^{(k)}-\frac{1}{2})\\
(p_1^{(k)}-\frac{1}{2})+i(p_2^{(k)}-\frac{1}{2})&1-p_3^{(k)}\\
\end{array}\right),\quad k=1,2,3,4
\end{equation}
where $p_1^{(k)},\,p_2^{(k)},\,p_3^{(k)}$ are probabilities of spin-$1/2$ projections equal to $+1/2$ on axes $\vec x,\,\vec y,\,\vec z,$ respectively for all four artificial qubits. Comparing matrix elements of matrices (\ref{eq.4.3}) and (\ref{eq.4.4}) with matrix elements of the same matrices given in the form (\ref{eq.4.5}) we obtain the following relations for diagonal matrix elements
\begin{eqnarray}
&&\rho_{33}=1-p_3^{(1)},\nonumber\\
&&\rho_{22}=1-p_3^{(2)}=p_3^{(4)},\label{eq.4.6}\\
&&\rho_{11}=p_3^{(1)}+p_3^{(2)}-1\nonumber
\end{eqnarray}
and for offdiagonal matrix elements
\begin{eqnarray}
&&\rho_{21}=(p_1^{(2)}-\frac{1}{2})+i(p_2^{(2)}-\frac{1}{2}),\nonumber\\
&&\rho_{31}=(p_1^{(1)}-\frac{1}{2})+i(p_2^{(1)}-\frac{1}{2}),\label{eq.4.7}\\
&&\rho_{32}=(p_1^{(4)}-\frac{1}{2})+i(p_2^{(4)}-\frac{1}{2}). \nonumber
\end{eqnarray}
One has the equalities
\begin{equation}\label{eq.4.8}
p_1^{(1)}=p_1^{(3)},\,p_2^{(1)}=p_2^{(3)},\,p_3^{(4)}=1-p_3^{(2)},\,p_3^{(1)}=p_3^{(3)}+p_3^{(4)}.
\end{equation}
For four qubits we have 12 parameters but the given equalities mean that 8 probabilities determine the state (\ref{eq.4.1}). Thus the density $3\times3$-matrix  $\rho$ (\ref{eq.4.1}) can be written in the form
\begin{equation}\label{eq.4.9}
\rho=\left(\begin{array}{ccc}
p_3^{(1)}+p_3^{(2)}-1\,\,\,&(p_1^{(2)}-\frac{1}{2})-i(p_2^{(2)}-\frac{1}{2})\,\,\,&(p_1^{(1)}-\frac{1}{2})-i(p_2^{(1)}-\frac{1}{2})\\
(p_1^{(2)}-\frac{1}{2})+i(p_2^{(2)}-\frac{1}{2})\,\,&1-p_3^{(2)}\,\,&(p_1^{(4)}-\frac{1}{2})-i(p_2^{(4)}-\frac{1}{2})\\
(p_1^{(1)}-\frac{1}{2})+i(p_2^{(1)}-\frac{1}{2})\,\,\,&(p_1^{(4)}-\frac{1}{2})+i(p_2^{(4)}-\frac{1}{2})\,\,&1-p_3^{(1)}\\
\end{array}\right)
\end{equation}
Here the probabilities $p_j^{(k)}$, $j=1,2,3,\,k=1,2,3,4$ satisfy the inequalities (\ref{eq.3.5}). Thus the inequality for one qubit state density matrix (\ref{eq.3.5}) is valid for all artificial qubit states density matrices associated with qutrit state (\ref{eq.4.9}).  Also the inequality $\det \rho\geq0$ provides the relation where cubic polynomial constructed from the probabilities $p_j^{(k)}$ must take nonnegative values.

In order to generalise the result obtained for the qutrit state we introduce the following notations
\[
p_3^{(1)}=p_3^{(33)},\quad p_3^{(2)}=p_3^{(22)}, \quad p_1^{(1)}=p_1^{(31)},\quad p_2^{(1)}=p_2^{(31)},\]
\[p_1^{(2)}=p_1^{(21)},\quad p_2^{(2)}=p_2^{(21)},\quad p_1^{(4)}=p_1^{(32)},\quad p_2^{(4)}=p_2^{(32)}.\]
Using the introduced notations we can rewrite the matrix elements of density matrix (\ref{eq.4.9}) in the form
\begin{eqnarray}
&&\rho_{j k}=(p_1^{(j k)}-\frac{1}{2})+i(p_2^{(j k)}-\frac{1}{2}),\quad j>k\nonumber\\
&&\rho_{j j}=1-p_3^{(j\,j)},\quad j\geq 2,\nonumber\\
&&\rho_{11}=1-\sum_{j=2}^3 \rho_{j j}.\label{eq.4.10.a}
\end{eqnarray}
This form is preserved if we consider arbitrary qudit density $N\times N$-matrix $\rho_{j k}$, $j,k=1,2,\ldots, N.$ In this case we have the artificial qubit probabilities $p_{1,2}^{(j k)}$, $j>k$; $p_3^{(j\,j)},\, j\geq2$. For example, qutrit density matrix reads
\begin{equation}\label{eq.4.11}
\rho=\left(\begin{array}{ccc}
p_3^{(33)}+p_3^{(22)}-1\,\,\,&(p_1^{(21)}-\frac{1}{2})-i(p_2^{(21)}-\frac{1}{2})\,\,\,&(p_1^{(31)}-\frac{1}{2})-i(p_2^{(31)}-\frac{1}{2})\\
(p_1^{(21)}-\frac{1}{2})+i(p_2^{(21)}-\frac{1}{2})\,\,&1-p_3^{(22)}\,\,&(p_1^{(32)}-\frac{1}{2})-i(p_2^{(32)}-\frac{1}{2})\\
(p_1^{(31)}-\frac{1}{2})+i(p_2^{(31)}-\frac{1}{2})\,\,\,&(p_1^{(32)}-\frac{1}{2})+i(p_2^{(32)}-\frac{1}{2})\,\,&1-p_3^{(33)}\\
\end{array}\right)
\end{equation}
It is worthy to note that in \cite{CherJRLR1} the qutrit density matrix was expressed in terms of probabilities but there was used different artificial qubit state.

The proof of the expression of the arbitrary density $N\times N$ - matrix in terms of probabilities $p_{1,2}^{(j k)}$, $p_3^{(j,j)}$ where $j>k$ describing spin-$1/2$ projections on axes $\vec x,$ $\vec y,$ $\vec z$, respectively can be done using the induction method. We illustrate this statement considering the case of $N=4$. The density matrix $\rho_{j k}$, $j.k=1,2,3,4$ can be embedded into $6\times 6$-density matrices,
\begin{equation}\label{eq.4.11.a}
\rho(1)=\left(\begin{array}{cc}
\rho&0\\
0&0\\
\end{array}\right),\quad
\rho(2)=\left(\begin{array}{cc}
0&0\\
0&\rho\\
\end{array}\right).
\end{equation}
Considering these matrices as density matrices of qutrit--qubit systems and applying the partial tracing procedure we get four density matrices $R(1),\,R(2),\,R(3),\,R(4)$ of the form
\begin{eqnarray}
&&R(1)=\left(\begin{array}{cc}
\rho_{11}+\rho_{22}+\rho_{33}&\,\,\rho_{14}\\
\rho_{41}&\,\rho_{44}\\
\end{array}\right), \quad
R(2)=\left(\begin{array}{ccc}
\rho_{11}+\rho_{44}&\,\,\rho_{12}&\,\,\rho_{13}\\
\rho_{21}&\,\,\rho_{22}&\,\,\rho_{23}\\
\rho_{31}&\,\,\rho_{32}&\,\,\rho_{34}\\
\end{array}\right),\nonumber\\
&&\label{eq.4.11.b}\\
&&R(3)=\left(\begin{array}{cc}
\rho_{11}&\,\,\rho_{14}\\
\rho_{41}&\,\rho_{22}+\rho_{33}+\rho_{44}\\
\end{array}\right), \quad
R(4)=\left(\begin{array}{ccc}
\rho_{22}&\,\,\rho_{23}&\,\,\rho_{24}\\
\rho_{32}&\,\,\rho_{33}&\,\,\rho_{34}\\
\rho_{42}&\,\,\rho_{43}&\,\,\rho_{44}+\rho_{11}\\
\end{array}\right).\nonumber
\end{eqnarray}
Since we know the expressions for the qubit states $R(1),\,R(3)$ and qutrit states $R(2),\,R(4)$ in terms of probabilities we can express the density matrix $\rho_{j k}$, $j,k=1,2,3,4$ in terms of these probabilities. We have  shown that the transition from qubit density matrix (\ref{eq.3.3}) to qutrit density matrix (\ref{eq.4.1})
gives the expression of the qutrit density matrix in the form (\ref{eq.4.11}). The analogous procedure is used to consider the $4\times4$-matrix and it repeats all the steps of the procedure to consider the $3\times3$-matrix and only adds extra matrix elements. Thus we have the density matrix of the ququart state in the form
\[\rho=\quad\quad\quad\quad\quad\,\]
\[
\small{\left(\begin{array}{cccc}
p_3^{(44)}+p_3^{(22)}+p_3^{(33)}-2&p_1^{(21)}-\frac{1}{2}-i(p_2^{(21)}-\frac{1}{2})
&p_1^{(31)}-\frac{1}{2}-i(p_2^{(31)}-\frac{1}{2})&p_1^{(41)}-\frac{1}{2}-i(p_2^{(41)}-\frac{1}{2})\\
p_1^{(21)}-\frac{1}{2}+i(p_2^{(21)}-\frac{1}{2})&1-p_3^{(22)}
&p_1^{(32)}-\frac{1}{2}-i(p_2^{(32)}-\frac{1}{2})&p_1^{(42)}-\frac{1}{2}-i(p_2^{(42)}-\frac{1}{2})\\
p_1^{(31)}-\frac{1}{2}+i(p_2^{(31)}-\frac{1}{2})&p_1^{(32)}-\frac{1}{2}+i(p_2^{(32)}-\frac{1}{2})
&1-p_3^{(33)}&p_1^{(43)}-\frac{1}{2}-i(p_2^{(43)}-\frac{1}{2})\\
p_1^{(41)}-\frac{1}{2}+i(p_2^{(41)}-\frac{1}{2})&p_1^{(42)}-\frac{1}{2}+i(p_2^{(42)}-\frac{1}{2})&
p_1^{(43)}-\frac{1}{2}+i(p_2^{(43)}-\frac{1}{2})\,&1-p_3^{(44)}\\
\end{array}\right).}\]
\begin{eqnarray}&&\label{eq.4.12}
\end{eqnarray}
Since the numbers $p_{1,2,3}^{(j k)}$ are probabilities the inequalities for all density matrices hold
\begin{equation}\label{eq.4.13}
\mbox{Re}\,\rho_{j k}+\frac{1}{2}\geq0, \quad \mbox{Im}\,\rho_{j k}\leq\frac{1}{2}.
\end{equation}
Also for arbitrary qudit density $N\times N$-matrix one has new entropic inequality
\begin{equation}\label{eq.4.14}
\left(\frac{1}{2}-\mbox{Im}\,\rho_{j k}\right)
\ln\left[\frac{\left(\frac{1}{2}-\mbox{Im}\,\rho_{j k}\right)}{\left(\frac{1}{2}-\mbox{Im}\,\rho_{j' k'}\right)}\right]+
\left(\frac{1}{2}+\mbox{Im}\,\rho_{j k}\right)
\ln\left[\frac{\left(\frac{1}{2}+\mbox{Im}\,\rho_{j k}\right)}{\left(\frac{1}{2}+\mbox{Im}\,\rho_{j' k'}\right)}\right]\geq0
\end{equation}
Here $j\neq k,\,j'\neq k',\,i,k,j',k'=1,2,\ldots,N$. Another new entropic inequality for arbitrary qudit state reads
\begin{equation}\label{eq.4.15}
\rho_{j j}\ln\left[\frac{\rho_{j j }}{\left(\frac{1}{2}\mp\mbox{Im}\,\rho_{j' k}\right)}\right]+
(1-\rho_{j j})\ln\left[\frac{(1-\rho_{j j })}{\left(\frac{1}{2}\pm\mbox{Im}\,\rho_{j' k}\right)}\right]\geq0.
\end{equation}
The entropic inequality of the form
\begin{equation}\label{eq.4.16}
\ln 2\geq -\left(\frac{1}{2}\mp\mbox{Im}\rho_{j k}\right)\ln\left(\frac{1}{2}\mp\mbox{Im}\rho_{j k}\right)-
\left(\frac{1}{2}\pm\mbox{Im}\rho_{j k}\right)\ln\left(\frac{1}{2}\pm\mbox{Im}\rho_{j k}\right)\geq0
\end{equation}
holds for arbitrary density matrix of qudit state.

\section{Triada of Malevich's squares and quantum suprematism picture of qudit states}
The discussed probability representation of qudit states can be pictorially illustrated \cite{CherJRLR1,CherJRLR2,CherJRLR32017} by using Malevich's squares known in suprematism direction in art \cite{Malevich}.  It means that for arbitrary probabilities $p^{(j k)}_1$, $p^{(j k)}_2$, $p^{(j k)}_3$ determined by matrix elements $\rho_{ j k}$ of density $N\times N$ - matrix $\rho$ one can introduce the triangles with three sides 
\[l_{ j k}^{(s)}=(2+2(p_s^{(j k)})^2-4p_s^{(j k)}-2p_{s+1}^{(j k)}+2(p_{s+1}^{(j k)})^2+2p_s^{(j k)}p_{s+1}^{(j k)})^{1/2},\quad  s=1,2,3. \]
Three squares (red, black and white) called "triada of Malevich's squares'' have the areas  
\[S_{j k}^{(s)}=\left[l_{j k}^{(s)}\right]^2, \quad s=1,2,3.\]
There is the bijective map of the density matrix of the qudit states with the matrix elements $\rho_{j k}$ and the set of the triadas of Malevich's squares.  Due to nonnegativity condition for density matrix $\rho_{j k}$ the areas of the Malevich's squares satisfy the inequalities described by the inequalities for polynomials expressed in terms of the probabilities $p^{(j k)}_s$, $s=1,2,3.$ The number of parameters determining the density matrix $\rho_{j  k}$, $j,k=1,2,\ldots,N$ equals $N^2-1$. This number provides the number of triadas of Malevich's squares describing the qudit states. For example $N=3$  (qutrit) considered in \cite{CherJRLR32017} one has three triadas of Malevich's squares with extra constraints for the sides of the squares. In Fig.1 we give a pictorial description of qudit state with N=4 (spin 3/2 state) using a mosaic constructed from five triadas of Malevich's squares. The ququart density matrix is mapped onto this ''quantum suprematism'' picture on the plane.

\begin{figure}
\includegraphics[width=150mm]{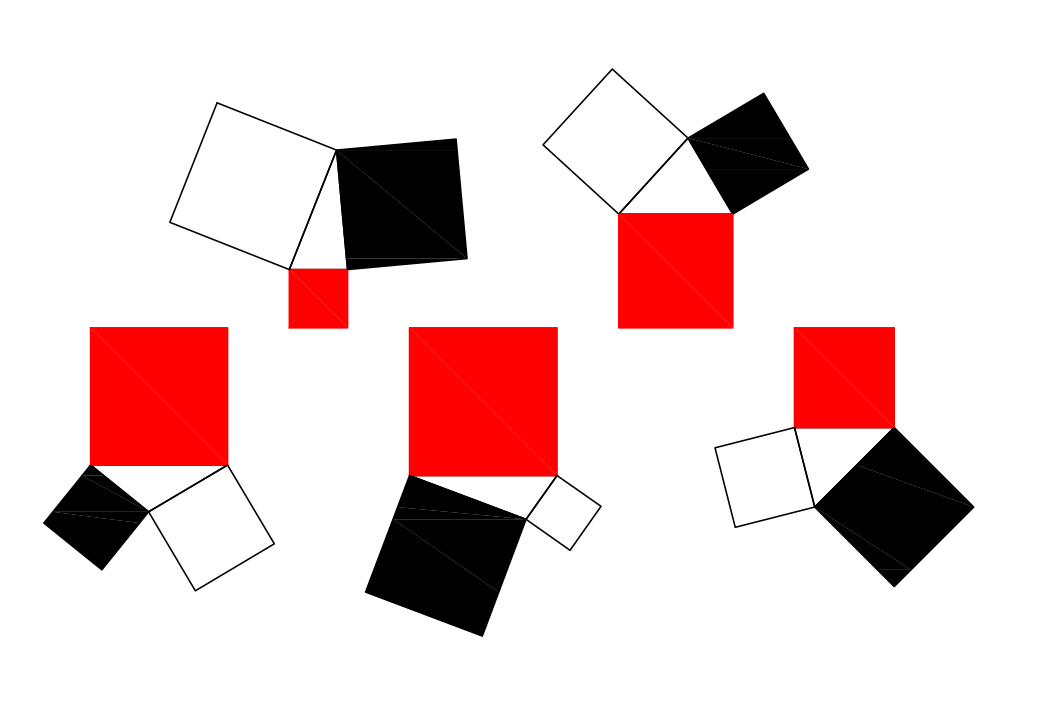}\\
\caption{''Quantum suprematism'' picture on the plane: five triadas of Malevich's squares associated with density matrix of ququart state.}
\end{figure}

Each square in five triadas has the area determined by the matrix elements of the spin $3/2$ density matrix  (\ref{eq.4.12}) expressed in terms of the probabilities $p_s^{(j k)}$. Thus five triadas of Malevich's squares illlustrate a new parametrization of the density matrix by the $15$ independent probabilities. There are quantum correlations in the geometric picture of the squares corresponding to nonnegativity condition of the density matrix. These conditions are expressed as nonnegativity conditions for eigenvalues of the matrix  (\ref{eq.4.12}).  

One can also map the probabilities $p^{(j k)}_{s}$ onto the parameters $X_{j k}^{(s)}$ of Bloch ball describing the artificial qubits in view of the relation $2 p^{(j k)}_{s}-1=X_{j k}^{(s)}$. These parameters satisfy the inequalities $\sum_{s=1}^3\left(X_{j k}^{(s)}\right)^2\leq1$ as well as other quantum constraints associated with nonnegativity of the density matrix. One can introduce the entropic characteristic of the qudit state as the sum of entropies of small artificial qubits. For example, for qubit (N=2) the entropy reads
\[H=-p_1\ln p_1-(1-p_1)\ln(1-p_1)-p_2\ln p_2-(1-p_2)\ln(1-p_2)-p_3\ln p_3-(1-p_3)\ln(1-p_3).\]
The relation of this entropy with von--Neumann entropy $S$ can be clarified if one takes into account the expression of entropy $S$ with the probabilities $p_1,$ $p_2$, $p_3$, i.e.
\begin{eqnarray*}
S=&&-\left(\frac{1}{2}+\sqrt{\sum_{k=1}^3(p_k-1/2)^2}\right)\ln\left(\frac{1}{2}+\sqrt{\sum_{k=1}^3(p_k-1/2)^2}\right)\nonumber\\
&&-\left(\frac{1}{2}-\sqrt{\sum_{k=1}^3(p_k-1/2)^2}\right)\ln\left(\frac{1}{2}-\sqrt{\sum_{k=1}^3(p_k-1/2)^2}\right).\nonumber
\end{eqnarray*}
Another entropic characteristics of  the qudit state is connected with areas of Malevich's squares. The sum $\Sigma$ of the areas of all the Malevich's squares determining the qudit state provides the probability characteristic of the state
\[{\cal P}_{j k}^{(s)}=\frac{\sum_{j k}^{(s)}}{\Sigma},
\]
where $\sum_{j k}^{(s)}$ is the area of one square. The Shannon entropy
\[H_{\Sigma}=-\sum_{j k}\sum_{(s)}{\cal P}_{j k}^{(s)}\ln{\cal P}_{j k}^{(s)}
\] 
gives the characteristics associated with the properties of the qudit state.

\section{Conclusion}
To resume we point out the main results of our work. We show that an arbitrary state of the qudit (spin-$j$ state, $N$-level atom state) can be described by $N^2-1$ probability distributions of artificial classical variables. These variables can be identified with variables associated with classical coins. It means that arbitrary $N\times N$ - density matrix of $N$-level system contains offdiagonal matrix elements of the form
$p^{(j k)}_{1}-\frac{1}{2}-i\left(p^{(j k)}_{2}-\frac{1}{2}\right)$, $j<k$, where the nonnegative numbers $p^{(j k)}_{1}$, $1-p^{(j k)}_{1}$ and $p^{(j k)}_{2},$ $1-p^{(j k)}_{2}$ are probability distributions describing the classical coin position "up" and "down". These probabilities correspond to $j\, k$-th spin-projection $+1/2$ on direction $x$ and direction $y$, respectively. The diagonal matrix elements $\rho_{j j}$, where $1<j\leq N$ have the form $1-p^{(j j)}_{3}$.
Thus the density matrix is mapped (see Eq.(\ref{eq.4.10.a})) onto $N^2-1$ probabilities which can be interpreted as probabilities from distributions associated with random positions of $N^2-1$ classical coins. It is the main result of our work. The density matrix can be demonstrated on the plane pictorially as the set of triadas of Malevich's squares which have been introduced in the framework of quantum suprematism picture of the qubit and qutrit states in \cite{CherJRLR1}-\cite{CherJRLR32017}. New entropy--information inequalities are obtained for matrix elements of the density matrix using the relation of the matrix elements with probability distributions.
Another result of our work is considering the parametric oscillator in probability representation. We obtain the new entropic relations for Frank--Condon factors. The vibronic structure of the electronic lines in polyatomic molecules is measured experimentally. Since this structure is determined by the Franck-Condon factors the correlations in the factors correspond to the correlations in the line intensities. Consequently the values of Shannon information given, e.g. by  Eq.(\ref{eq.2.7}) reflect the correlations in vibronic structure of electronic lines in polyatomic spectra. The deeper clarification of the relation of the dependence of Shannon information value to the correlations in the polyatomic molecule spectra will be considered in future publication. It is worthy to note that the results known in molecular spectroscopy can be related to properties of systems used in quantum optics and information (see, e.g. \cite{Huh1,Huh3}). These relations (inequalities) can be checked in experiments with superconducting circuits. In \cite{CherJRLR2} it was found for qubit state that the quantum observables (arbitrary Hermitian matrices) are mapped onto sets of classical random variables. This map can be constructed for arbitrary qudit observables.

Thus the quantum statistics formalism for arbitrary system can be mapped onto classical statistics formalism of the sets of classical coins and corresponding random variables. This problem will be considered in future publication.

\subsection*{Acknowledgments}

V. I. M. thanks Professor D. Schuch and Professor M. Ramek for invitation to participate  in Symmetries in Science Symposium, Bregenz, 2017,  and for the hospitality. We thank P. I. Lysikhin, artist for help with mosaic presentation of Malevich's squares. 



\begin{thebibliography}{99}

\bibitem{Dirac}  Dirac P 1930 {\it The Principles of Quantum Mechanics}, Oxford University Press


\bibitem{Schr26} Schr\"{o}dinger E 1926 {\it Ann. Phys.} {\bf 79}, 361; {\bf 81}, 109



\bibitem{Landau27} Landau L D 1927 {\it Z. Phys.} {\bf 45} 430



\bibitem{vonNeumann27}von~Neumann J 1927 {\it G\"ottingenische Nachrichten} {\bf 11} 245


\bibitem{Wigner32} Wigner E 1932 {\it Phys. Rev.} {\bf 40} 749

\bibitem{Glauber63} Glauber R J  1963 {\it Phys. Rev. Lett.}  {\bf 10} 84 

\bibitem{Sud63} Sudarshan E C C 1963 {\it Phys. Rev. Lett} {\bf 10} 277 

\bibitem{Cah} Cahill K E and Glauber R J 1969 {\it Phys. Rev.} {\bf 177} 1882 

\bibitem{Husimi} Husimi K 1940 {\it Proc. Phys. Math. Soc. Jpn} {\bf 23} 264 

\bibitem{Kano56} Kano Y 1965 {\it J. Math. Phys.} {\bf 6} 1913 

\bibitem{Stratonovich57}  Stratonovich R L 1957 {\it J. Exp. Theor. Phys.} {\bf 5} 1206

\bibitem{TombesPLA96} Mancini S  Man'ko V  I and Tombesi P 1996 {\it Phys. Lett. A} {\bf
213} 1


\bibitem{TombesiFoundPhyus96} Mancini S Man'ko V I and Tombesi P 1997 {\it Found. Phys.}  {\bf 27}  801 

\bibitem{JRLR1997} Man'ko O V and Man'ko V I 1997 {\it J. Russ. Las. Res.} {\bf 18} 407

\bibitem{Raymer93}Smithey D T, Beck M, Raymer M G and Faridani A 1993 {\it Phys. Rev.
Lett.} {\bf 70} 1244

\bibitem{BerBer} Bertrand J and Bertrand P 1989 {\it Found. Phys.}  {\bf 17} 397 


\bibitem{VogRis} Vogel K and Risken H 1989 {\it  Phys. Rev. A} {\bf 40}, 2847 


\bibitem{BentyYacolb}Man'ko M A and Man'ko V  I 2012  ``Tomographic entropic inequalities
in the probability representation of quantum mechanics,'' in: R.~Bijker, (Ed.), {\it Beauty in Physics: Theory and Experiment: in Honor of Francesco Iachello on the Occasion of His 70th Birthday, Hacienda Cocoyoc, Mexico, 14-18 May 2012}, AIP Conference Proceedings, Vol.~1488, pp.~110--121



\bibitem{Radon1917} Radon J 1917  {\it Berichte Sr\"{a}chsische Akademie der Wissenschaften} {\bf 29}, 262 


\bibitem{DodPLA} Dodonov V V and  Man'ko V I 1997 {\it Phys. Lett. A} {\bf 229} 335 



\bibitem{OlgaJETP}Man'ko V I and Man'ko O V 1997 {\it J. Exp. Theor. Phys.} {\bf 85} 430


\bibitem{Bregence} Man'ko O V 1998 "Tomography of spin states and classical formulation of quantum mechanics," in:  B.~Gruber and M.~Ramek~(Eds.), {\it Proceedings of
International Conference ``Symmetries in Science X'' (Bregenz, Austria, 1997)}, Plenum Press, New York , p.~207.


\bibitem{Weigert1} Weigert S 2000 {\it Phys. Rev. Lett.}  {\bf 84} 802 

\bibitem{Weigert2}  Amiet J P and Weigert S 1999 {\it J.~Opt.~B: Quantum Semiclass. Opt.} {\bf 1}  L5 

\bibitem{PainiDariano} D'Ariano G M Maccone L and Paini M 2003 {\it J.~Opt.~B: Quantum Semiclass. Opt.} {\bf  5} 77 

\bibitem{IbortPhysScr2009} Ibort A Man'ko V I Marmo G Simoni A and Ventriglia F 2009 {\it  Phys. Scr.} {\bf 79}  065013 

\bibitem{IbortPhysScr2015}  Asorey M  Ibort A Marmo G and Ventriglia F 2015 {\it  Phys. Scr.} {\bf 90}  074031

\bibitem{MendesJRusLasRes}  Man'ko M A Man'ko V I and  Mendes R V 2006 {\it J. Russ. Las. Res.} {\bf 27} 507

\bibitem{MAVI} Man'ko M A Man'ko V I 2018 {\it J. Russ. Las. Res.} {\bf 39} 1

\bibitem{MalkinMolSP1076} Doktorov E V  Malkin I A and Man'ko V I 1975 {\it J. Mol. Spectroscop.} {\bf 56} 1 

\bibitem{MalkinMolSpec1977}  Doktorov E V Malkin I A and Man'ko V I 1977 {\it J. Mol. Spectroscop.} {\bf  64} 302 


\bibitem{HuberHu} Huh J and Berger R 2012  {\it J. Phys.: Conf.  Ser.} {\bf 380} 1 

\bibitem{ZhabrakJRusLasRes} Zhebrak E D 2016 {\it J. Rus. Las. Res.}  {\bf 37} 123

\bibitem{MarmoJPA2016}  Man'ko V I  Marmo G  Ventriglia F and Vitale P 2017 {\it J. Phys. A: Math. and Theor.} {\bf 50} 335302 

\bibitem{CherJRLR1} Chernega V N  Man'ko O V and Man'ko V I 2017 {\it J. Rus. Las. Res.} {\bf 38} 1 


\bibitem{CherJRLR2} Chernega V N  Man'ko O V and Man'ko V I 2017 {\it J. Rus. Las. Res.} {\bf 38} 324 


\bibitem{CherJRLR32017} Chernega V N Man'ko O V and Man'ko V I 2017 {\it J. Russ. Las. Res.} {\bf 38} 141 

\bibitem{Malevich} Aleksandra Shatskikh, {\it Black Square: Malevich and the Origin of Suprematism},
Yale University Press, New Haven (2012).

\bibitem{TombesiSemOpt} Mancini S Man'ko V I and Tombesi P. 1995 {\it J.~Opt.~B: Quantum Semiclass. Opt.} {\bf 7} 615


\bibitem{MendesPLA} Man'ko M A  Man'ko V I and  Mendes R V 2001 {\it J. Phys. A: Math. and Gen.} {\bf 34} 8321 

\bibitem{MalkinMankoPhysLet1966}  Malkin I A  Man'ko V I and  Trifonov D A 1969 {\it Phys. Let. A} {\bf 30} 414 


\bibitem{Ermakov} Ermakov V P  1880 {\it Universitetskie Izvestiya}, Kiev,  {\bf No. 9} 1  [in Russian]

\bibitem{183} Dodonov V V and  Man'ko V I 1987  in: M. A. Markov (Ed.), {\it "Invariants and the Evolution of Nonstationary Quantum Systems", Proceedings of Lebedev Physics Institute}, Vol. 183
~Nauka, Moscow, p. 182


\bibitem{Huh1} Shen Y Y  Huh J J  Lu Y Y  Zhang J J  Zhang K K  Zhang S S and  Kim K K  2018 {\it Chem. Sci.} {\bf 9} 836


\bibitem{Huh2} Huh J  Guerreschi G G  Peropadre B  McClean J R and Aspuru-Guzik A  2015  {\it Nature Photon.} {\bf 9} 615 


\bibitem{Huh3}   Huh J and  Yung M H 2017 {\it Sci. Rep} {\bf 7} 7462


\bibitem{Shannon} Shannon C E 1948 {\it Bell. Syst. Tech. J.} {\bf 27} 379; 623

\bibitem{SeilovJRLR}  Man'ko V I and  Seilov Z 2017 {\it J. Russ. Las. Res.}  {\bf 38} 50 

\bibitem{SeilovJRLR1} Man'ko V I and Seilov Z 2016 {\it J. Russ. Las. Res.}  {\bf 37} 591 


\bibitem{Vaxjo} Man'ko O V 2012  ``Quantum tomography of current and voltage states in nanoelectri circuits,''  in: M. D.'Ariano, S.-M. Feri, E. Haven, B. Hiesmayer, G. Jaeger, A. Khrennikov and J.-A. Larsson, (Ed.), {\it Foundation of Probabilility and Physics-6, Vaxjo, Sweden, 14-16 June 2011}, AIP Conference Proceedings, Vol.~1424, pp.~221--233

\bibitem{physicascr2013} Man'ko O V 2013 {\it Phys. Scr.}  {\bf T153} 014046

\bibitem{kiktenko}Fedorov A K Kiktenko E O Man'ko O V and Man'ko V I 2015 {\it Phys. Rev. A} {\bf 91} 042312

\bibitem{LewisReisenfeld} Lewis H R  and Riesenfeld W B 1969 {\it J.~Math.~Phys.} {\bf 10} 1458 


\bibitem{Shuch1} Cruz H  Schuch D  Castanos O and  Rosas-Ortiz O 2015 {\it Annals of Physics} {\bf 360} 44 

\bibitem{Shuch2} Schuch D 2014 {\it J. Phys.: Conf. Ser.} {\bf 538} 012019 

\bibitem{Shuch3} Schuch D  Guerrero J  Lopez-Ruiz F F and Aldaya V 2015 {\it Phys. Scr.}  {\bf 90}  045209 

\bibitem{ShuchCastanos}  Rosas-Ortiz O  Castanos O and  Schuch D 2015 {\it J. Phys. A: Math. Theor.} {\bf 48} 445302 


\end{thebibliography}
\end{document}